# Cycle-Constrained Adversarial Denoising Convolutional Network for PET Image Denoising: Multi-Dimensional Validation on Large Datasets with Reader Study and Real Low-Dose Data

Yucun Hou, Fenglin Zhan, Xin Cheng, Chenxi Li, Ziquan Yuan, Runze Liao, Haihao Wang, Jianlang Hua, Jing Wu, Jianyong Jiang

*Abstract*—Positron emission tomography (PET) is a critical tool for diagnosing tumors and neurological disorders but poses radiation risks to patients, particularly to sensitive populations. While reducing injected radiation dose mitigates this risk, it often compromises image quality. To reconstruct full-dose-quality images from low-dose scans, we propose a Cycle-constrained Adversarial Denoising Convolutional Network (Cycle-DCN). This model integrates a noise predictor, two discriminators, and a consistency network, and is optimized using a combination of supervised loss, adversarial loss, cycle consistency loss, identity loss, and neighboring Structural Similarity Index (SSIM) loss. Experiments were conducted on a large dataset consisting of raw PET brain data from 1,224 patients, acquired using a Siemens Biograph Vision PET/CT scanner. Each patient underwent a 120-seconds brain scan. To simulate low-dose PET conditions, images were reconstructed from shortened scan durations of 30, 12, and 5 seconds, corresponding to 1/4, 1/10, and 1/24 of the full-dose acquisition, respectively, using a custom-developed GPU-based image reconstruction software. The results show that Cycle-DCN significantly improves average Peak Signal-to-Noise Ratio (PSNR), SSIM, and Normalized Root Mean Square Error (NRMSE) across three dose levels, with improvements of up to 56%, 35%, and 71%, respectively. Additionally, it achieves contrast-to-noise ratio (CNR) and Edge Preservation Index (EPI) values that closely align with full-dose images, effectively preserving image details, tumor shape, and contrast, while resolving issues with blurred edges. The results of reader studies indicated that the images restored by Cycle-DCN consistently received the highest ratings from nuclear medicine physicians, highlighting their strong clinical relevance.

*Index Terms*—Positron emission tomography (PET), Image Denoising, Cycle-Consistent Adversarial Networks.

This work was supported in part by the National Natural Science Foundation of China (No. 12475336 & 12105018), Beijing Nova Program (No. Z211100002121129 & No. 20230484413). (Yucun Hou and Fenglin Zhan are co-first authors. Corresponding author: Jianyong Jiang.)

Yucun Hou, Chenxi Li, Ziquan Yuan, Runze Liao, Haihao Wang, Jianlang Hua, Jing Wu and Jianyong Jiang are with School of Physics and Astronomy, Beijing Normal University, Beijing 100875. China and with Key Laboratory of Beam Technology of Ministry of Education, Beijing Normal University, Beijing 100875. China (e-mail: hou9021@163.com; 2929204456@qq.com; 1365953243@qq.com; liaorunze@nuaa.edu.cn; 374089798@qq.com; jianlang_hua@163.com; jingwu@bnu.edu.cn; jianyong@bnu.edu.cn).

Fenglin Zhan is with Department of Nuclear Medicine，The First Affiliated Hospital of USTC, Division of Life Sciences and Medicine, University of Science and Technology of China, Hefei, Anhui, 230001, China；Wuxi School of Medicine, Jiangnan University, No. 1800, Lihu Avenue, Wuxi, 214000, China (e-mail: zhan209@qq.com ).

Xin Cheng is with Department of Nuclear Medicine (PET-CT Center), National Cancer Center/National Clinical Research Center for Cancer/Cancer Hospital, Chinese Academy of Medical Sciences and Peking Union Medical College, Beijing, 100021, China (e-mail: dr_chengxin@cicams.ac.cn ).

## I. INTRODUCTION

Positron emission tomography (PET) is a non-invasive imaging technique that uses radiotracers to visualize metabolic and physiological processes[1-5]. After injection, the radiotracer undergoes β+ decay, emitting a positron, which annihilates with an electron, producing two nearly opposite gamma-rays. These are detected as coincidence events, providing insight into *in vivo* functions[6-9]. However, the radioactive nature of the tracers poses a significant radiation hazard, particularly to sensitive populations, including infants, children, and adolescents[10, 11]. Minimizing radiation exposure in PET is essential for mitigating the associated risks of radioactive tracers. Low-dose PET scanning presents a promising strategy for reducing the patient's radiation burden and is particularly relevant in maternal–fetal medicine, where such scans may facilitate safer investigations of fetal and placental physiology, serving as a less invasive alternative to traditional procedures.

Nevertheless, low-dose PET images are susceptible to increased noise levels, which considerably degrade image quality. As the radiation dose decreases, noise becomes more pronounced, obscuring tissue and lesion structures and thereby limiting the diagnostic utility of the images[11]. Consequently, methods that accurately reconstruct full-dose-quality images from low-dose scans are of considerable clinical importance, as they can reconcile the necessity for reduced radiation exposure with the demand for accurate and reliable diagnostic information. Additionally, in regions with a high demand for PET scans, patients may face delays between the injection of the radiotracer and the subsequent imaging due to scheduling constraints. This waiting period allows for radioactive decay, which naturally reduces the effective dose by the time imaging occurs. Such decay can adversely affect imaging quality and diagnostic accuracy, particularly in the detection of small lesions. This issue underscores the clinical importance of accurately restoring full-dose PET images from low-dose scans.

Several convolutional neural network (CNN) models have been developed for low-dose PET image denoising, aiming to restore low-dose images to full-dose quality. Jiang et al. introduced a semi-supervised model that applies region normalization to pre-segmented tissues and enforces structural consistency to generate standard-dose PET images from unpaired low-dose PET data[12]. Jang et al. proposed a spatial and channel-wise encoder-decoder transformer that integrates both spatial and channel information for enhanced PET image denoising[13]. Wang et al. compared five advanced deep learning models—U-Net[14], Enhanced Deep Super-Resolution Network (EDSR)[15], Generative Adversarial Network (GAN)[16], Swin Transformer Network (SwinIR)[17], and Vision Transformer (ViT)[18] with an EDSR encoder—across six count levels[19], concluding that SwinIR and U-Net substantially improved diagnostic image quality. However, most of models prioritize noise reduction at the cost of critical details, such as edges and small structures, often leading to overly smooth images that resemble traditional filters, thereby losing essential information[20, 21]. To address this limitation, Liu et al. proposed a blending strategy that combines outputs from models trained on both high- and low-noise images, balancing noise reduction with image clarity[21]. In medical image denoising, one of the central challenges lies in restoring structural details, particularly the shape and activity of small lesions, necessitating methods that reduce noise while preserving diagnostic integrity.

CycleGAN[22] was designed to enable image-to-image translation without paired images, allowing for the modification of specific features while retaining essential image characteristics. It employs two generators to learn mappings between domains and two discriminators to differentiate real from generated images. The model uses three key losses: adversarial loss to align the distribution of generated images with the target domain, cycle consistency loss to ensure input images can be reconstructed from their translated versions, and identity loss to preserve images when processed by the generator corresponding to their original domain. These losses allow CycleGAN to perform tasks like style transfer, object transfiguration, and image enhancement. Zhou et al. applied CycleGAN to PET image denoising, treating low-dose and full-dose images as separate domains[23]. By incorporating adversarial, cycle-consistency, identity, and supervised losses, they achieved noise reduction while maintaining lesion contrast. However, as a generative model, CycleGAN introduces uncertainty during training, making it difficult to preserve small tissue structures and tumor contours, requiring careful design of the loss function to constrain the model effectively.

In this study, we proposed and extensively investigated a novel Cycle-constrained Adversarial Denoising Convolutional Network (Cycle-DCN) aimed at reducing noise while preserving diagnostic integrity in PET images. Specifically, the contributions and innovations of this work include the following.

1. We developed Cycle-DCN, a model that performs domain mapping by extracting noise through a noise extractor while incorporating a consistency network to gather auxiliary information from neighboring slices. Cycle-DCN leverages a combination of supervised loss, adversarial loss, cycle-consistency loss, identity loss, and neighboring SSIM loss. The model dynamically adjusts the weight of these losses during training, and we systematically evaluated the impact of each component and loss function on the quality of denoised images.

2. A total of 1,224 raw datasets were acquired from a clinical Siemens Biograph PET/CT scanner and reconstructed using custom-developed QuanTOF[24] algorithm. Utilizing complex brain PET images as the experimental dataset, we rigorously assessed Cycle-DCN's capability to preserve fine details in the denoised images.

3. A multi-dimensional evaluation framework was employed to assess denoising performance, including metrics such as Peak Signal-to-Noise Ratio (PSNR), Structural Similarity Index (SSIM), Normalized Root Mean Square Error (NRMSE), Contrast-to-Noise Ratio (CNR), Edge Preservation Index (EPI), and the Hausdorff Distance of edge contours. Additionally, six nuclear medicine physicians (each with over 10 years of experience) from various hospitals evaluated and scored the denoised images. This comprehensive evaluation validates the model's effectiveness in noise reduction, preservation of brain sulci and gyri, and restoration of tumor shape and contrast.

## II. MATERIALS AND METHODS

### A. PET Image Reconstruction

In this study, 1,224 head scan datasets were acquired using a Siemens Biograph Vision PET/CT scanner. The raw list-mode datasets were reconstructed using QuanTOF[24], a GPU-accelerated, Bayesian penalized likelihood reconstruction algorithm. QuanTOF integrates time-of-flight (TOF) information and complete correction techniques, ensuring high-quality PET image reconstruction. To simulate low-dose PET conditions, images were reconstructed from scan shortened durations of 30, 12, and 5 seconds, corresponding to 1/4, 1/10, and 1/24 of the full dose, respectively, with 120 seconds of scan data serving as the full-dose reference. As illustrated in Fig. 1, the reconstructed images via QuanTOF using list-mode datasets, due to accurate modeling of imaging physics, exhibit clearer details and sharper edges compared to those reconstructed via the vendor's e7-tools solution using sinogram datasets, providing a more reliable assessment of the model's ability to preserve fine details.

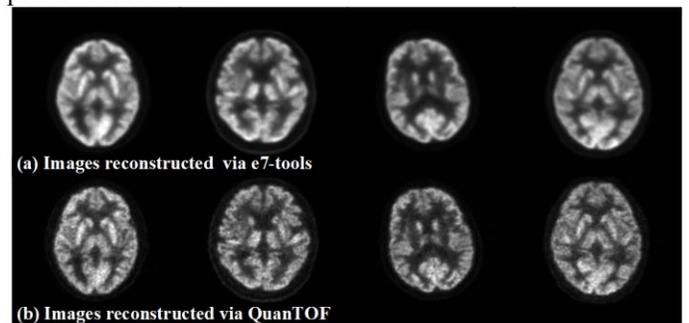

Fig. 1. Images reconstructed by e7-tools (top row) and QuanTOF (bottom row).

### B. Cycle-constrained

Fig. 2 presents a detailed diagram of the proposed Cycle-DCN architecture, which replaces the two generators in CycleGAN with a noise predictor. This noise predictor extracts noise from low-dose images to facilitate their transformation into full-dose images. The noise-extracted image is designated as the 'fake' image, while the unprocessed image is designated as the 'real' image. Consequently, a real low-dose image is transformed into a fake full-dose image after noise removal, and a real full-dose image becomes a fake low-dose image after noise addition, as illustrated in Fig. 2(b) and (c). The structural details of the noise predictor are provided in Fig. 3(a).

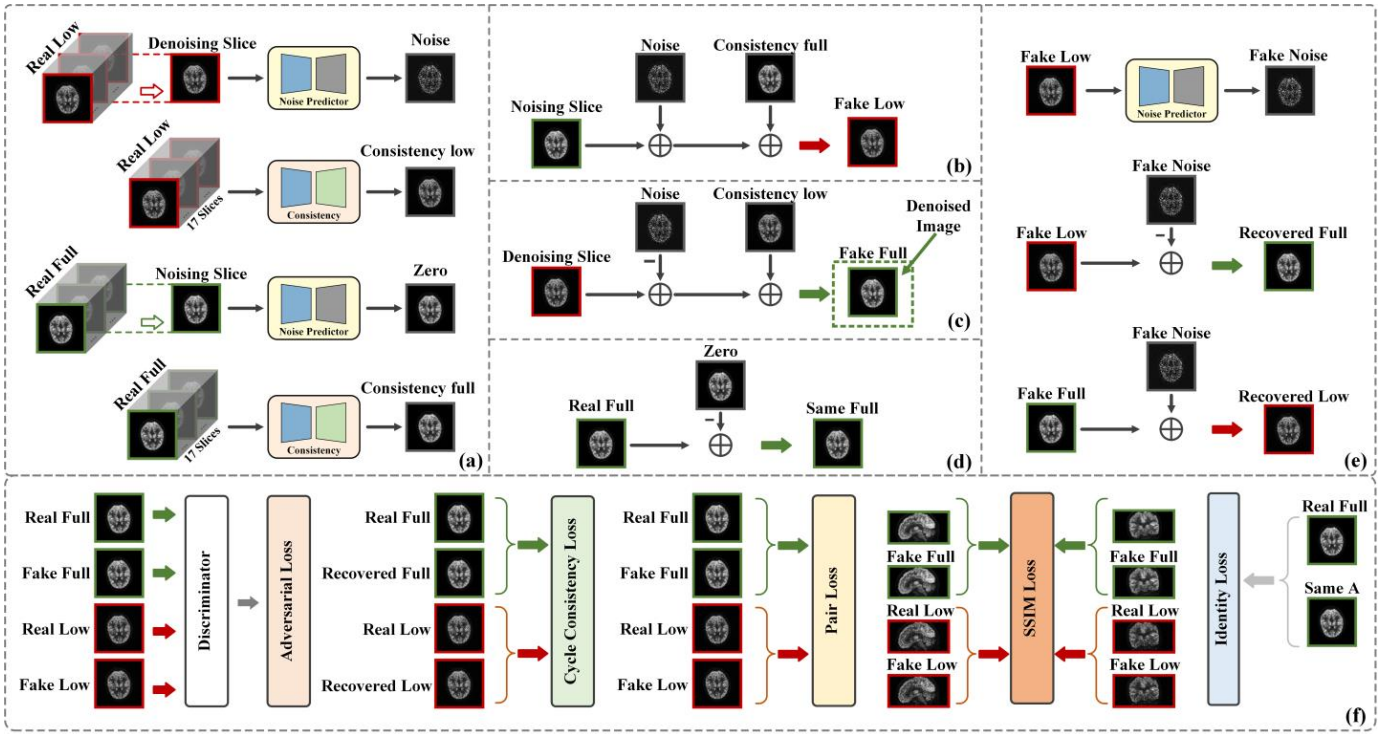

Fig. 2. Diagrams of the Cycle-DCN model. (a) Noise extraction process; (b), (c) Noise removal and addition processes; (d), (e) Generation of the same and recovered images; (f) Composition of the Cycle-DCN loss function.

Similar to CycleGAN, the proposed Cycle-DCN employs two discriminators that evaluate whether the images are real or fake, contributing to the adversarial loss. This loss encourages the model to generate images that resemble real ones. The architecture of the two discriminators is illustrated in Fig. 3(b). The full-dose and low-dose domains are represented as $u \sim p_{data}(u)$ and $o \sim p_{data}(o)$, respectively. The denoising process is denoted as $P$, and noise addition as $P'$. The full-dose discriminator is referred to as $D_U$ and the low-dose discriminator as $D_O$. The adversarial loss, $\mathcal{L}_{\text{GAN}}$, is calculated as:

$$\mathcal{L}_{\text{GAN}}(P, D_O, D_U, O, U) = \mathbb{E}_{u \sim p_{data}(u)}\left[\left(D_O(P'(u)) - 1\right)^2\right]$$
$$+ \mathbb{E}_{o \sim p_{data}(o)}\left[\left(D_U(P(o)) - 1\right)^2\right] \quad (1)$$

Here, $P(o)$ represents the fake full-dose image, and $P'(u)$ represents the fake low-dose image. The objective of the noise predictor is to minimize $\mathcal{L}_{\text{GAN}}$, encouraging the two discriminators to classify the fake images as 'True' while the discriminators aim to distinguish them as 'False'. The loss functions for $D_O$ and $D_U$ are defined as:

$$\mathcal{L}_{\text{GAN}}(P, \boldsymbol{D_O}, O, U) = \mathbb{E}_{u \sim p_{data}(u)}[D_O(P'(u))^2]$$
$$+ \mathbb{E}_{o \sim p_{data}(o)}[(D_O(o) - 1)^2] \quad (2)$$

$$\mathcal{L}_{\text{GAN}}(P, \boldsymbol{D_U}, O, U) = \mathbb{E}_{u \sim p_{data}(u)}[(D_U(u) - 1)^2]$$
$$+ \mathbb{E}_{o \sim p_{data}(o)}[(D_U(P(o)))^2] \quad (3)$$

The model performs a cyclic process where noise is extracted from the input low-dose image to generate a fake full-dose image via the denoising process P. This fake full-dose image is then used to recover a low-dose image through the noise addition process $P'$, as illustrated in Fig. 2(e). If the denoising is effective, the 'recovered' image will closely match the real image. The cycle consistency loss is calculated by measuring the difference between these images, guaranteeing that the denoising process reliably eliminates noise without compromising other essential image details. It is calculated as:

$$\mathcal{L}_{\text{CYC}}(P, O, U) = \mathbb{E}_{u \sim p_{data}(u)}\left[\left\|P(P'(u)) - u\right\|_1\right]$$
$$+ \mathbb{E}_{o \sim p_{data}(o)}\left[\left\|P'(P(o)) - o\right\|_1\right] \quad (4)$$

Additionally, when a full-dose image is input, the model avoids any modifications to prevent over-denoising. As illustrated in Fig. 2(a) and (d), the noise extracted from the real full-dose image should be zero, ensuring the image remains unchanged after the denoising process $P$, producing an identical 'same' full-dose image. This is enforced through identity mapping loss, which measures the difference between the real and the same full-dose images as:

$$\mathcal{L}_{\text{Identity}}(P, O, U) = \mathbb{E}_{u \sim p_{data}(u)}[\|P(u) - u\|_2] \quad (5)$$

Finally, supervised loss is incorporated based on a substantial dataset of paired low-dose and full-dose images, which is defined as:

$$\mathcal{L}_{\text{Sup}}(P, O, U) = \mathbb{E}_{u \sim p_{data}(u)}[\|P'(u) - o\|_2]$$
$$+ \mathbb{E}_{o \sim p_{data}(o)}[\|P(o) - u\|_2] \quad (6)$$

### C. Weight

In adversarial training, the weights assigned to the four losses are critical for optimizing model performance. Each loss serves a distinct purpose, necessitating adjustments based on the input data distribution and the specific training stage. The overall loss function is expressed as:

$$\mathcal{L} = \lambda_G \mathcal{L}_{\text{GAN}} + \lambda_C \mathcal{L}_{\text{CYC}} + \lambda_I \mathcal{L}_{\text{Identity}} + \lambda_P \mathcal{L}_{\text{Sup}} \quad (7)$$

To maintain balanced contributions from all four losses, their weights are dynamically adjusted throughout training according to their current values, ensuring they remain within the same order of magnitude. To prevent division by zero errors and numerical instability resulting from minuscule values, a constant term $\varepsilon$ was incorporated into the denominator. The weight update mechanism is specified as:

$$\lambda_{total} = \frac{1}{\mathcal{L}_{\text{GAN}} + \varepsilon} + \frac{1}{\mathcal{L}_{\text{CYC}} + \varepsilon} + \frac{1}{\mathcal{L}_{\text{Identity}} + \varepsilon} + \frac{1}{\mathcal{L}_{\text{Sup}} + \varepsilon}$$

$$\lambda_G = \frac{1/\mathcal{L}_{\text{GAN}} + \varepsilon}{\lambda_{total}}, \quad \lambda_C = \frac{1/\mathcal{L}_{\text{CYC}} + \varepsilon}{\lambda_{total}} \quad (8)$$

$$\lambda_I = \frac{1/\mathcal{L}_{\text{Identity}} + \varepsilon}{\lambda_{total}}, \quad \lambda_P = \frac{1/\mathcal{L}_{\text{Sup}} + \varepsilon}{\lambda_{total}}$$

### D. Neighboring Slices Assisted and Sagittal SSIM Loss

In Cycle-DCN, each horizontal slice is denoised independently. However, given the continuous nature of brain tissue, neighboring slices often contain valuable structural information that may not be fully captured in the current slice. To leverage this continuity, we developed a consistency network that extracts structural information from neighboring slices and integrates it into the denoising process, as illustrated in Fig. 2(a), thereby enhancing the quality of the reconstructed images. The architecture of this model is presented in Fig. 3(a). This auxiliary data is incorporated into both the noise extraction and noise addition stages. Furthermore, to ensure consistent performance across different anatomical planes, the SSIM values in the sagittal and coronal planes are incorporated into the loss function as additional constraints.

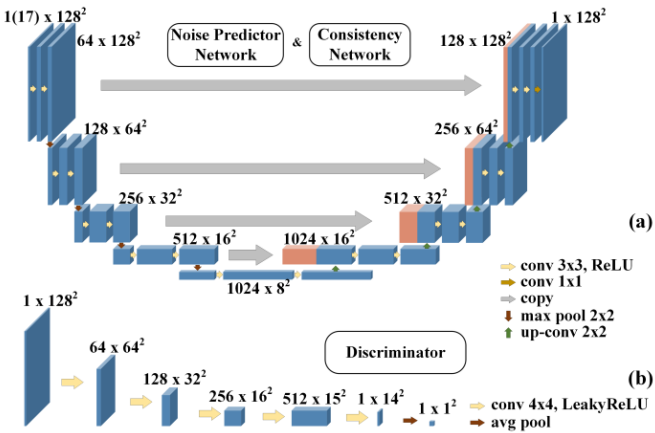

Fig. 3. Network architecture diagrams for the (a)Noise Predictor Network and Consistency Network; (b) Discriminator.

## III. EXPERIMENTS

### A. Dataset and Implementation

A total of 1,224 head scan datasets, acquired from a Siemens Biograph Vision PET/CT scanner between November 2023 and June 2024, were included in this study. The patients had an average age of 60 and an average weight of 62.8 kg. All patients underwent standard clinical whole-body PET scans followed by individual head scans with an injected dose of 278.7 ± 50.8 MBq of $^{18}$F-fluorodeoxyglucose ($^{18}$F-FDG). The datasets were partitioned into 1,024 cases for training, 100 for testing, and the remaining 100 for validation. All images were cropped to a size of 128 × 256 × 256, and min-max normalization was applied to scale the intensity values to a range of [0, 1].

The experiments were conducted on an Ubuntu 24.04 LTS system with an Intel Xeon Gold 6226R CPU, 256GB of RAM, and two NVIDIA GeForce RTX 4090 GPUs, each with 24GB of memory. Python was used for program development, with the PyTorch deep learning framework employed for model implementation.

### B. Evaluation Metrics

The model's performance was evaluated using several quantitative metrics, including Peak Signal-to-Noise Ratio (PSNR), Structural Similarity Index (SSIM), and Normalized Root Mean Square Error (NRMSE). These metrics assessed the overall quality of the denoised images in terms of noise reduction and structure preservation. In addition, the model's ability to accurately recover tumor uptake was specifically assessed using the Contrast-to-Noise Ratio (CNR)[13, 25], which measures the relative contrast between the tumor region and the surrounding background. The calculation of CNR is provided as:

$$CNR = \frac{|\mu_t - \mu_b|}{\sigma_b} \quad (9)$$

Here, $\mu_t$ and $\mu_b$ represent the mean intensities of the tumor and background regions, respectively, while $\sigma_b$ denotes the standard deviation of the background intensity.

To further assess the model's ability to preserve and restore edge details, the Edge Preservation Index (EPI)[26, 27] was employed. The EPI measures the ratio of gradients between the full-dose and denoised images, providing an indicator of how well the edge structures are maintained post-denoising. The EPI is calculated as:

$$EPI = \frac{\sum_{i,j} |\nabla I_{\text{denoised}}(i,j)|}{\sum_{i,j} |\nabla I_{\text{full}}(i,j)|} \quad (10)$$

Here, $\nabla I_{\text{full}}(i,j)$ and $\nabla I_{\text{denoised}}(i,j)$ represent the gradient maps of the full-dose and denoised images at pixel positions $i$ and $j$, respectively. To quantify the similarity of edge structures between the denoised and full-dose images, edge extraction was conducted using the Canny operator[28]. The extracted edges were then compared by calculating the Hausdorff Distance (HD)[29], which measures the maximum distance between the sets of edge points in the two images. The formula for the Hausdorff Distance, where a smaller HD indicates closer alignment of the edge structures between the full-dose and denoised images, is provided as:

$$H(A,B) = \max \begin{Bmatrix} \sup_{a \in A} \inf_{b \in B} d(a,b), \\ \sup_{b \in B} \inf_{a \in A} d(b,a) \end{Bmatrix} \quad (11)$$

In Equation (11), $A$ and $B$ represent the sets of contour points extracted from the full-dose and denoised images, respectively. For each point $a$ in set $A$, the minimum distance $\inf_{b \in B} d(a,b)$ to the nearest point in set B is calculated, the maximum value $\sup_{a \in A} \inf_{b \in B} d(a,b)$ among these minimum distances is then determined. Similarly, $\sup_{b \in B} \inf_{a \in A} d(b,a)$ represents the maximum of the minimum distances from points in $B$ to the nearest points in $A$. The larger of these two maximum values is taken as the Hausdorff Distance.

### C. Reader Study

To evaluate the denoising performance from a clinical perspective, we conducted a reader study involving six nuclear medicine physicians, each with a minimum of 10 years of experience. The physicians were affiliated with various institutions: The First Affiliated Hospital of University of Science and Technology of China (two physicians), Peking Union Medical College Hospital (one), Cancer Hospital Chinese Academy of Medical Sciences (one), Beijing Friendship Hospital (one), and Shanghai Pulmonary Hospital (one).

Physicians scored the five model images by comparing their similarity to full-dose images, focusing on factors such as edge detail preservation, restoration in clinically relevant areas, blur, noise levels, image definition, contrast, and overall visual quality. The scoring criteria were as follows: 5 points for nearly identical to the full-dose image, 4 points for very similar, 3 points for some differences, 2 points for significant differences, and 1 point for substantial differences.

The evaluation was performed in a double-blind format, focusing on denoised images generated by five different models. For each low-dose version, 30 cases were randomly selected from the test dataset. Each physician was presented with seven images (five denoised images from the models, one low-dose image, and one full-dose image) from each patient simultaneously. Custom software was developed to facilitate the convenient comparison of the images.

## IV. RESULT

### A. Quantitative evaluations

#### 1) PSNR, SSIM, and NRMSE Comparison

The proposed Cycle-DCN model was compared with established methods, including widely used U-Net[14], classic denoising model DnCNN[30], the more recent image restoration model CGNet[31], and CycleGAN. Table I provides the quantitative results, showing the mean, standard deviation, and statistical significance based on paired t-tests for each evaluation metric. Figure 4(a) presents a visual comparison of a representative slice across the five models, along with the corresponding full-dose and low-dose images at three different dose levels.

For 1/4 low-dose images, the PSNR, SSIM, and NRMSE metrics are similar across models. While DnCNN slightly outperforms Cycle-DCN in some measures, these differences are not statistically significant ($p > 0.05$). However, for the 1/24 low-dose images, Cycle-DCN achieves superior metric values. As shown in Fig. 4(a), U-Net and DnCNN result in excessive smoothing, leading to increased deviation from the full-dose images at lower dose levels, with small sulci becoming nearly indistinguishable. In contrast, Cycle-DCN remains closer to the full-dose images, demonstrating better robustness.

In the enlarged views in Fig. 4(b), Cycle-DCN uniquely preserves brain structures such as sulci and gyri, producing visual outputs more consistent with full-dose images compared to other models.

#### 2) Tumor Restoring Performance

Figures 5(a) and 5(b) show denoised tumor images at three low-dose levels of two representative cases. For the 1/4 low-dose images, all five models effectively denoise the images while preserving the tumor shape. However, as the dose decreases (1/10 and 1/24 low-dose), noise causes the tumor regions to blur or become nearly invisible in the denoised images of all models except Cycle-DCN.

In contrast, Cycle-DCN consistently restores tumor shape even at lower dose levels. To quantitatively assess the ability to restore tumor contrast, the CNR was calculated for the denoised 1/4 low-dose tumor images. Tumor and background regions were selected using a threshold segmentation method, encompassing the entire tumor in the axial direction, as shown in Fig. 5(c) for nine cases that containing tumors, where the red mask represents the tumor and the blue mask the background region. The CNRs of the nine cases are reported in Table II.

As expected, full-dose images achieved the highest CNR, indicating superior tumor detectability. Among the five models, Cycle-DCN achieved the highest CNR in 8 out of 9 denoised images and the second-highest in the remaining image. This demonstrates its superior ability to restore tumor contrast, closely matching the performance of full-dose images.

TABLE I
QUANTITATIVE COMPARISON OF MEAN PSNR, SSIM AND NRMSE AMONG DIFFERENT MODELS ON THE TEST DATASET

| Method | PSNR (p-value) | SSIM (p-value) | NRMSE (p-value) |
|---|---|---|---|
| 1/4 Low-dose | 21.430 ±1.870 | 0.727 ±0.031 | 0.813 ±0.026 |
| U-Net | 33.458 ±1.981(0.304) | 0.933 ±0.017(<0.001) | 0.213 ±0.055(0.351) |
| DnCNN | **33.671 ±2.064(0.062)** | 0.935 ±0.018(0.111) | **0.208 ±0.057(0.051)** |
| CGNet | 33.508 ±1.964(0.993) | 0.935 ±0.017(<0.05) | 0.213 ±0.058(0.509) |
| Cycle-GAN | 31.304 ±1.706(<0.001) | 0.895 ±0.020(<0.001) | 0.277 ±0.051(<0.001) |
| Cycle-DCN | **33.507 ±1.994** | **0.936 ±0.016** | **0.212 ±0.050** |
| 1/10 Low-dose | 20.074 ±1.794 | 0.657 ±0.031 | 0.950 ±0.012 |
| U-Net | 31.284 ±2.004(0.625) | 0.902 ±0.022(<0.05) | 0.276 ±0.084(0.586) |
| DnCNN | **31.321 ±1.944(0.255)** | **0.907 ±0.022(<0.001)** | **0.275 ±0.085(0.982)** |
| CGNet | 29.813 ±1.698(<0.001) | 0.893 ±0.022(<0.001) | 0.325 ±0.094(<0.001) |
| Cycle-GAN | 29.573 ±1.759(<0.001) | 0.873 ±0.023(<0.001) | 0.333 ±0.077(<0.001) |
| Cycle-DCN | **31.253 ±2.012** | **0.903 ±0.022** | **0.275 ±0.071** |
| 1/24 Low-dose | 19.723 ±1.763 | 0.639 ±0.031 | 0.989 ±0.004 |
| U-Net | 30.211 ±1.750(0.370) | 0.877 ±0.022(<0.001) | 0.313 ±0.101(0.106) |
| DnCNN | 29.888 ±1.620(<0.001) | 0.879 ±0.021(<0.001) | 0.326 ±0.105(<0.001) |
| CGNet | 29.165 ±1.612(<0.001) | 0.880 ±0.023(<0.05) | 0.353 ±0.112(<0.001) |
| Cycle-GAN | 28.697 ±1.272(<0.001) | 0.860 ±0.021(<0.001) | 0.381 ±0.113(<0.001) |
| Cycle-DCN | **30.256 ±1.829** | **0.882 ±0.022** | **0.310 ±0.094** |

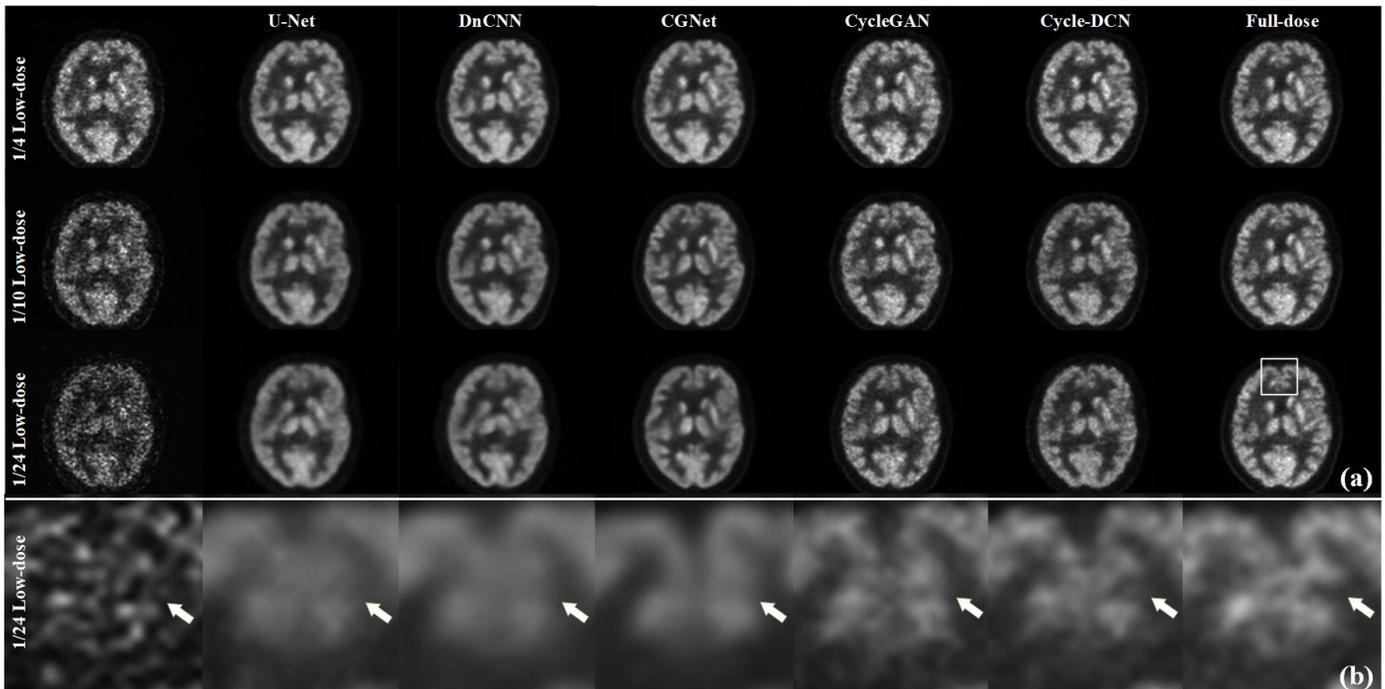

Fig. 4. (a) Denoised images from various models at three different low-dose levels for a representative case; (b) The partially enlarged sections of 1/24 low-dose images, showcasing detailed visual comparisons across models.

TABLE II
COMPARISON OF CNR ACROSS 9 CASES FOR DIFFERENT MODELS

| Method | Case-1 | Case-2 | Case-3 | Case-4 | Case-5 | Case-6 | Case-7 | Case-8 | Case-9 |
|---|---|---|---|---|---|---|---|---|---|
| Full-dose | 6.5 | 6.27 | 11.734 | 7.223 | 6.153 | 13.529 | 4.165 | 9.819 | 9.575 |
| 1/4 Low-dose | 4.247 | 3.963 | 7.904 | 4.719 | 3.921 | 7.655 | 2.373 | 5.358 | 5.022 |
| Unet | 4.478 | 4.061 | **9.611** | 5.399 | 4.677 | 8.645 | 2.617 | 6.679 | 5.785 |
| DnCNN | 4.402 | 4.47 | 7.701 | 4.996 | 3.854 | 7.579 | 2.529 | 6.442 | 5.919 |
| CGNet | 3.715 | 4.069 | 8.187 | 4.914 | 3.591 | 6.385 | 2.453 | 5.597 | 5.157 |
| CycleGAN | 3.378 | 3.593 | 4.243 | 4.226 | 3.126 | 5.977 | 2.249 | 5.302 | 4.241 |
| Cycle-DCN | **4.58** | **4.644** | 9.451 | **5.724** | **5.642** | 8.882 | **2.711** | 6.708 | **6.766** |

### 3) Edge Restoring Performance

The human brain features complex sulcal and gyral structures, making the clarity and accuracy of these contours essential for clinical diagnosis. To assess the models' effectiveness in preserving these edge details, the EPI was calculated between the denoised images and the full-dose images. As presented in Table III, Cycle-DCN demonstrates EPI values closer to 1, signifying that its total image gradient is more similar to that of the full-dose image, thus indicating superior edge preservation.

To visually emphasize the edge contours, Canny edge detection was applied to the denoised images, as displayed in Fig. 5(d) for one representative case. Low-dose images, due to noise, tend to exhibit more false edges, while U-Net, DnCNN and CGNet produce fewer extracted edges as a result of excessive smoothing. In contrast, the edges in Cycle-DCN and CycleGAN-denoised images appear visually closer to the full-dose images. To quantitatively assess this similarity, the Hausdorff distance was computed between the edges of the full-dose and denoised images. As shown in Table III, Cycle-DCN achieves the smallest Hausdorff distance, indicating superior edge preservation.

TABLE III
COMPARISON OF EPI AND HAUSDORFF DISTANCE FOR DIFFERENT MODELS (Value ±Std)

| Method | EPI | Hausdorff Distance |
|---|---|---|
| 1/4 Low-dose | 1.204 ±0.146 | 24.493 ±3.580 |
| Unet | 0.934 ±0.105 | 20.190 ±3.636 |
| DnCNN | 0.891 ±0.090 | 21.266 ±3.946 |
| CGNet | 0.911 ±0.103 | 21.491 ±3.830 |
| CycleGAN | 1.066 ±0.118 | 17.889 ±3.875 |
| Cycle-DCN | **0.994 ±0.108** | **16.764 ±3.648** |

### B. Reader study

The results of the reader study are summarized in Fig. 6. For the entire dataset of 30 patients acquired using the Siemens Biograph Vision PET/CT scanner, all six readers consistently preferred the images denoised by our proposed Cycle-DCN model. This preference led to significantly higher average scores for Cycle-DCN, highlighting its effectiveness in producing high-quality PET image.

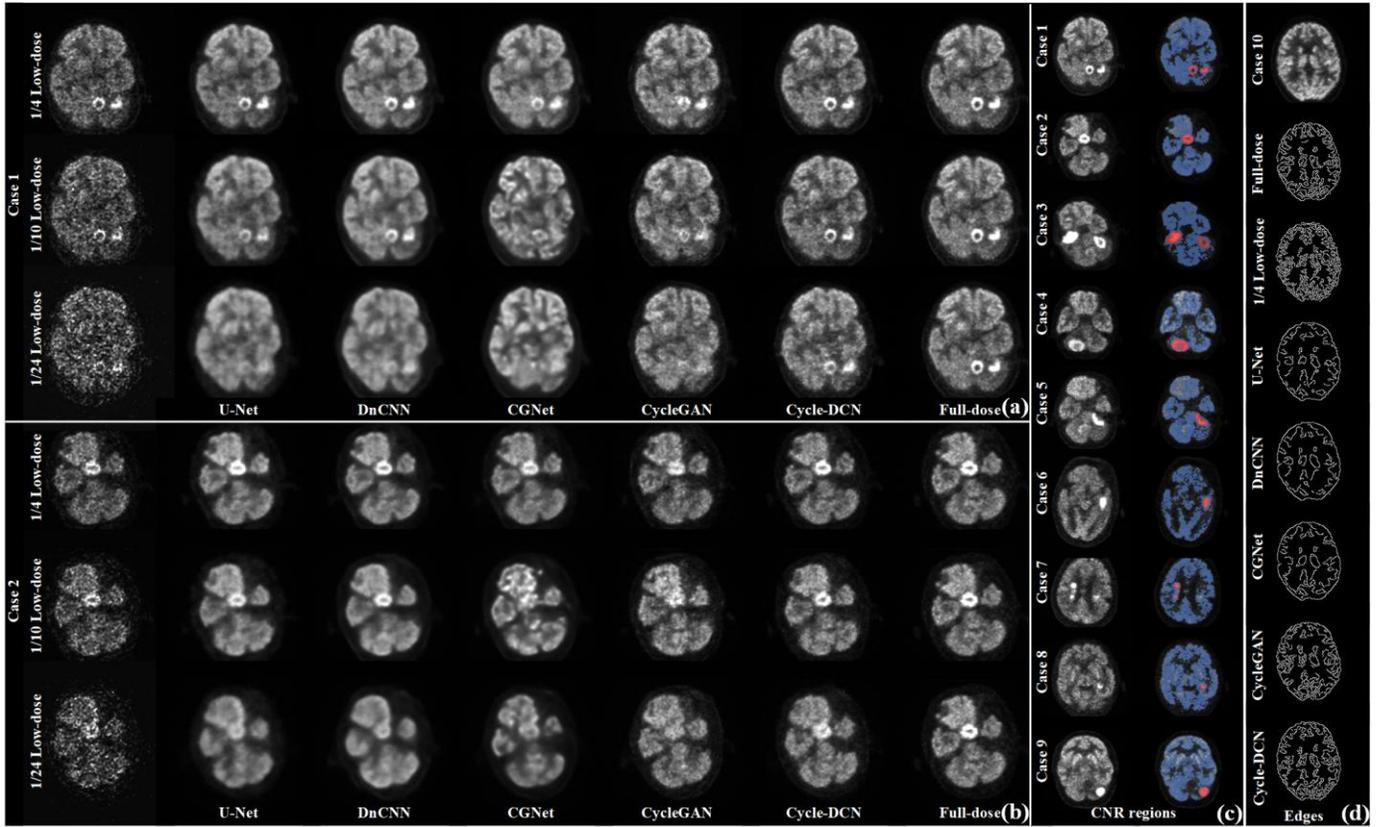

Fig. 5. (a), (b) Denoised images from various models at three different low-dose levels for two representative cases; (c) The selected tumor and background regions; (d) The edge-extracted images obtained using the Canny operator.

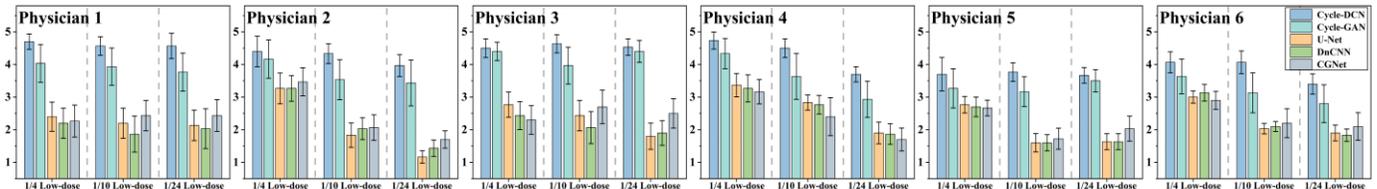

Fig. 6. The scoring results from the six physicians for different models on the randomly selected 30 cases in the test dataset.

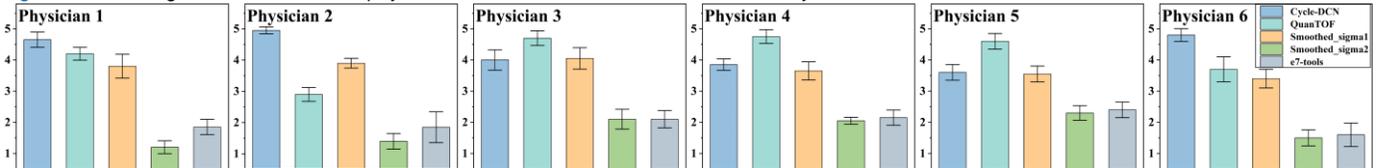

Fig. 7. The scoring results for real low-dose images processed using different algorithms on the randomly selected 20 cases in the test dataset.

## C. Real Low-dose images

In a clinical setting, several patients may receive tracer injections simultaneously, and PET scans are conducted sequentially after a waiting period of ~40 minutes. This protocol results in some patients being scanned at a lower dose at the time of scanning. The pre-trained 1/4 low-dose denoising model was applied to these cases, and its denoised images were compared with several alternatives: original reconstructed images, original reconstructed images processed with Gaussian filtering (σ = 1 or 2), and images reconstructed using the vendor's e7-tools, as shown in Fig. 8 for one representative case.

The original reconstructed images displayed noticeable noise, affecting image quality. Gaussian filtering (σ = 2) caused excessive smoothing, leading to significant blurring, a problem also observed in vendor-reconstructed images. Gaussian filtering (σ = 1) yielded images similar to those produced by our model, but with more noise, particularly around the edges, where our model showed clearer detail.

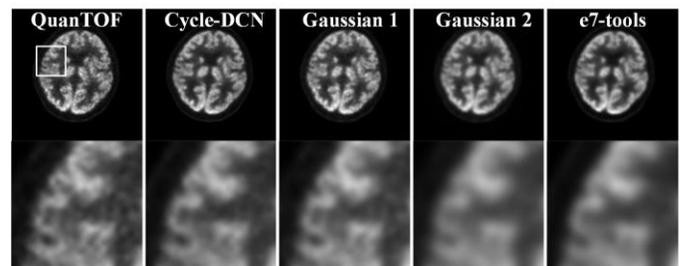

Fig. 8. Denoising results of real low-dose images using different methods (top row) and the enlarged sections (bottom row) for one representative case.

To evaluate the model's clinical relevance, the same six physicians were asked to score the images for suitability in clinical diagnosis based on a scoring system ranging from 1 to 5, where 5 indicates excellent image quality suitable for diagnosis, and 4 signifies good image quality, also acceptable for diagnostic purposes. A score of 3 reflects some defects in image quality, but these do not hinder diagnosis. A score of 2 indicates poor image quality that affects diagnosis, while a score of 1 means the image quality is unacceptable for diagnosis[32]. Twenty images were randomly selected for scoring, and the results are presented in Fig. 7. Notably, three physicians preferred the Cycle-DCN denoised images, while the other three favored the QuanTOF directly reconstructed images, despite the presence of some noise in the latter. All physicians unanimously assigned the lowest scores to images processed with Gaussian filtering (σ = 2) and reconstructed using e7-tools.

### D. Ablation Study

Cycle-DCN integrates several key loss functions: cycle consistency loss, adversarial loss, identity mapping loss, SSIM loss, and supervised loss. It also leverages information from neighboring slices to enhance noise reduction. To assess the contribution of these components, a series of ablation experiments were conducted to systematically evaluate their effectiveness.

*1) Impact of different loss:* To assess the effect of removing the supervised loss, we excluded it from the overall loss function. As shown in Table IV, this led to a decline in PSNR, SSIM, and NRMSE. Additionally, as illustrated in Fig. 9(a), the model without supervised loss introduces unintended structures in small regions. In contrast, the model with supervised loss generates images more closely resembling the full-dose images in these areas.

To evaluate the impact of the losses derived from GycleGAN, we removed all losses except the supervised loss. As shown in Table IV, the model with only supervised loss achieved slightly higher metrics for 1/4 low-dose images. However, for 1/24 low-dose images, it underperformed in restoring tumor shape and contrast compared to the full Cycle-DCN model with all losses included (Fig. 9b), and it produced smoother images, similar to U-Net and DnCNN (Fig. 5a).

TABLE IV
QUANTITATIVE COMPARISON OF MODELS WITH DIFFERENT LOSS AND COMPONENTS (Value ± Std)

| Method | PSNR | SSIM | NRMSE |
|---|---|---|---|
| 1/4 Low-dose | 21.430 ±1.870 | 0.727 ±0.031 | 0.813 ±0.026 |
| Without supervised loss | 32.568 ±2.140 | 0.922 ±0.019 | 0.236 ±0.053 |
| Only supervised loss | 33.939 ±1.979 | 0.941 ±0.016 | 0.202 ±0.052 |
| Without neighboring slices | 32.882 ±2.064 | 0.927 ±0.018 | 0.227 ±0.050 |
| Network-based fuse | 33.137 ±2.117 | 0.934 ±0.018 | 0.224 ±0.059 |
| Without SSIM loss | 33.420 ±2.055 | 0.935 ±0.016 | 0.214 ±0.051 |
| Cycle-DCN | 33.507 ±1.994 | 0.936 ±0.016 | 0.212 ±0.050 |

*2) Neighboring Slices Assisted:* Table IV presents a comparison of metrics with and without the neighboring slices information, showing a decline in performance when excluded. Fig. 9(c) demonstrates that without this information, denoised images exhibit noticeable transverse artifacts in the sagittal and coronal planes, reflecting insufficient continuity of brain tissue along the z-axis, which also reduces contrast.

Two fusion methods were tested: element-wise addition (in Cycle-DCN) and a network-based approach. Results indicate that the network-based fusion slightly underperforms, as shown in Table IV. Additionally, the impact of removing SSIM loss was assessed, and its absence also led to slightly lower performance metrics.

## V. DISCUSSION

The proposed Cycle-DCN model significantly enhances PSNR, SSIM, and NRMSE metrics, achieving improvements of 57%, 29%, and 74% on 1/4 low-dose images; 56%, 38%, and 74% on 1/10 low-dose images; and 54%, 38%, and 69% on 1/24 low-dose images, respectively. For 1/4 low-dose images, all tested models produced acceptable denoising results, but U-Net, DnCNN, and CGNet yielded marginally smoother images. In contrast, Cycle-DCN closely aligned with full-dose images, eff-

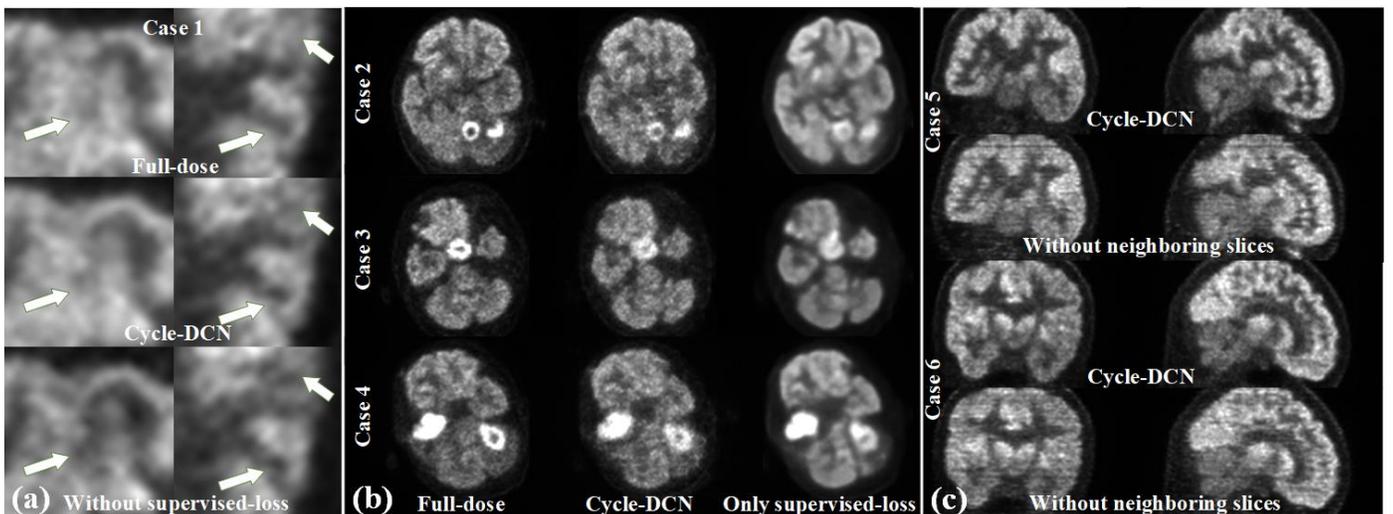

Fig. 9 (a) Comparison of the enlarged full-dose images, denoised images using Cycle-DCN, and denoised images from the model without supervised loss; (b) Comparison of the full-dose image, denoised images from Cycle-DCN, and denoised image from the model using only supervised loss for 3 different cases; (c) Comparison of denoised images in coronal (left column), and sagittal planes (right column) for two representative cases with/without neighboring slices assisted.

ectively preserving image details, tumor shape, and contrast, while resolving issues with blurred edges, as indicated by its EPI value closest to 1 and the smallest Hausdorff distance between edges detected by the Canny operator and those in full-dose images.

In terms of tumor visibility, all five models maintained distinguishable tumors in 1/4 low-dose images, with Cycle-DCN showing the highest CNR, indicating superior tumor contrast recovery. As the dose decreased to 1/10 and 1/24, U-Net, DnCNN, and CGNet produced increasingly over-smoothed images, rendering the tumors and many brain sulci indistinguishable, as shown in Fig. 5(a), (b) and the enlarged portions of Fig. 4(b). Conversely, Cycle-DCN more effectively preserved these structures, consistently recovering both shape and contrast of tumors even in heavily noise-contaminated 1/10 and 1/24 low-dose images, where tumor shapes were nearly unrecognizable compared to other models.

CycleGAN generates images that are visually similar to those produced by our model, primarily due to its incorporation of cycle consistency loss, adversarial loss, and identity mapping loss. However, as demonstrated in the denoised images presented in Figures 4 and 5, CycleGAN introduces erroneous structures, particularly at the 1/24 low-dose level, underscoring the inherent uncertainty in the outputs of generative models. The primary distinctions between our model, Cycle-DCN, and CycleGAN are twofold. First, Cycle-DCN does not rely on two-domain image translation; instead, it employs a noise extraction process to facilitate domain mapping, thereby simplifying the transformation process and sharpening the model's focus. Second, Cycle-DCN integrates supervised loss with a dynamic balance of additional losses, enhancing overall training performance.

In the reader study, images denoised using Cycle-DCN were consistently rated as more similar to full-dose images compared to those generated by U-Net, DnCNN, CGNet and CycleGAN, which demonstrated poorer fidelity. As illustrated in Figures 4 and 5, while U-Net, DnCNN, and CGNet effectively reduced noise in the 1/4 low-dose images, their denoised outputs often exhibited excessive smoothing, which intensified with decreasing dose levels, resulting in progressively blurrier images and lower ratings from physicians. In contrast, Cycle-DCN maintained clarity even at the 1/24 low-dose level, receiving consistently higher evaluations.

For real low-dose images processed using different methods, three physicians favored Cycle-DCN denoised images, while the other three preferred QuanTOF's direct reconstructions, despite the presence of some noise in the latter. All physicians assigned the lowest scores to images processed with Gaussian filtering (σ = 2) and reconstructed with e7-tools. As shown in Figure 8, Gaussian filtering (σ = 2) and e7-tools produced overly smoothed images, whereas Cycle-DCN preserved clearer contours and structural details while effectively reducing noise. Although Gaussian filtering (σ = 1) performed similarly to our model, it retained slightly more noise, leading to reduced clarity of structural edges. The preference for QuanTOF's reconstructions highlights the diagnostic value in structural details, though without corresponding full-dose images, it remains unclear whether these represents noise or clinically relevant features.

Achieving optimal results in Cycle-DCN requires dynamically adjusting the weights of various loss functions based on input data distribution, training stages, and outputs. The adversarial loss is derived from the discriminator's output, allowing the discriminator to distinguish between 'True' and 'False' classifications without fixed labels. This helps the model generate features from the target domain, compensating for data loss from lower injection doses. However, excessive adversarial loss weight can introduce artifacts, which is critical in medical imaging as it may compromise diagnostic accuracy. Thus, maintaining a moderate adversarial loss weight during training is essential. For supervised loss, paired full-dose and low-dose images serve as labels, akin to traditional supervised learning. This promotes faster convergence by guiding outputs toward full-dose images, while avoiding artifact generation. However, excessive supervised loss weight can lead to over-smoothing, resembling traditional denoising models, as demonstrated in the ablation study. Cycle consistency loss prevents unwanted changes to image content, preserving critical details such as edges, with specific constraints on the noise predictor to maintain edge clarity. Overemphasizing cycle consistency loss can result in zero noise output, rendering real and reconstructed images indistinguishable. Identity loss ensures that full-dose images remain unchanged and is scaled in proportion to the cycle consistency loss. For example, during the initial 0–100 epochs with 1/4 low-dose images, supervised loss should be weighted more heavily for faster convergence, followed by fine-tuning. For lower doses, such as 1/24, a higher adversarial loss weight is required, balanced by the supervised loss to avoid over-generation of structures. Proper adjustment of these loss weights is crucial to prevent negative effects from excessive weights and ensure precise denoising.

The proposed model exhibits several limitations that warrant further enhancement. First, our experiments have been limited to brain PET images. Future efforts will focus on developing denoising models specifically tailored for PET images from various organs, ultimately aiming to integrate these models into a comprehensive whole-body PET denoising framework. Second, the current dataset is exclusively derived from the Siemens Biograph Vision PET/CT scanner at the First Affiliated Hospital of University of Science and Technology of China. We plan to expand our data collection to include multiple hospitals and various scanner brands and models, targeting a dataset exceeding 10,000 cases. This expansion will facilitate large-scale, multicenter, and cross-device experimentation. Third, the PET scan data utilized in our experiments are based on patients injected with [18]F-FDG. Given that different radiotracers exhibit distinct absorption characteristics across tissues and organs, we aim to adapt and optimize our model for additional commonly used radiotracers, such as [18]F-ACBC and [68]Ga-DOTATATE, to further enhance its generalizability and robustness under diverse imaging conditions.

## VI. CONCLUSION

In this study, we present a Cycle-constrained Adversarial Denoising Convolutional Network (Cycle-DCN) designed to effectively reduce noise while preserving image clarity and ensuring the comprehensive restoration of fine structural features, such as brain sulci and the shape and activity of small lesions. The model was trained and tested on 1,224 datasets acquired using a Siemens Biograph Vision PET/CT scanner. Test results demonstrate that Cycle-DCN enhances the average PSNR, SSIM, and NRMSE values across three dose levels by up to 56%, 35%, and 71%, respectively. It successfully restores tumor shapes at all three low-dose levels and achieves tumor contrast that closely resembles full-dose images. Additionally, Cycle-DCN excels in accurately restoring the structures of brain sulci and gyri, closely mirroring those in full-dose images. The images produced exhibit an optimal Edge Preservation Index (EPI) and the smallest Hausdorff Distance for edges extracted via the Canny operator when compared to full-dose images. Furthermore, the reader study by six nuclear medicine physicians indicates that our model's denoised images are the most similar to full-dose images. The physicians expressed a preference for the Cycle-DCN denoised images over those processed with Gaussian filtering or vendor's e7-tools reconstruction, underscoring the strong clinical significance of our model.

While our proposed model has demonstrated favorable outcomes in brain PET image denoising, we are committed to expanding our data sources, adapting the model for various organs and radiotracers, and optimizing the model architecture to create a denoising solution applicable to whole-body PET imaging, thereby providing clearer and more accurate imaging support for clinical diagnosis.

## REFERENCES

[1] M. M. Ter-Pogossian, M. E. Phelps, E. J. Hoffman, and N. A. Mullani, "A positron-emission transaxial tomograph for nuclear imaging (PETT)," *Radiology,* vol. 114, no. 1, pp. 89-98, 1975.
[2] T. Beyer *et al.*, "A combined PET/CT scanner for clinical oncology," *Journal of nuclear medicine,* vol. 41, no. 8, pp. 1369-1379, 2000.
[3] J. W. Fletcher *et al.*, "Recommendations on the use of 18F-FDG PET in oncology," *Journal of Nuclear Medicine,* vol. 49, no. 3, pp. 480-508, 2008.
[4] E. M. Rohren, T. G. Turkington, and R. E. Coleman, "Clinical applications of PET in oncology," *Radiology,* vol. 231, no. 2, pp. 305-332, 2004.
[5] M. Schwaiger, S. Ziegler, and S. G. Nekolla, "PET/CT: challenge for nuclear cardiology," *Journal of Nuclear Medicine,* vol. 46, no. 10, pp. 1664-1678, 2005.
[6] E. Berg and S. R. Cherry, "Innovations in instrumentation for positron emission tomography," in *Seminars in nuclear medicine*, 2018, vol. 48, no. 4: Elsevier, pp. 311-331.
[7] S. R. Cherry, T. Jones, J. S. Karp, J. Qi, W. W. Moses, and R. D. Badawi, "Total-body PET: maximizing sensitivity to create new opportunities for clinical research and patient care," *Journal of Nuclear Medicine,* vol. 59, no. 1, pp. 3-12, 2018.
[8] G. El Fakhri, S. Surti, C. M. Trott, J. Scheuermann, and J. S. Karp, "Improvement in lesion detection with whole-body oncologic time-of-flight PET," *Journal of Nuclear Medicine,* vol. 52, no. 3, pp. 347-353, 2011.
[9] T. G. Turkington, "Introduction to PET instrumentation," *Journal of nuclear medicine technology,* vol. 29, no. 1, pp. 4-11, 2001.
[10] E. Robbins, "Radiation risks from imaging studies in children with cancer," *Pediatric blood & cancer,* vol. 51, no. 4, pp. 453-457, 2008.
[11] J. D. Schaefferkoetter, J. Yan, D. W. Townsend, and M. Conti, "Initial assessment of image quality for low-dose PET: evaluation of lesion detectability," *Physics in Medicine & Biology,* vol. 60, no. 14, p. 5543, 2015.
[12] C. Jiang, Y. Pan, Z. Cui, D. Nie, and D. Shen, "Semi-supervised standard-dose PET image generation via region-adaptive normalization and structural consistency constraint," *IEEE transactions on medical imaging,* vol. 42, no. 10, pp. 2974-2987, 2023.
[13] S.-I. Jang *et al.*, "Spach Transformer: Spatial and channel-wise transformer based on local and global self-attentions for PET image denoising," *IEEE transactions on medical imaging,* vol. 43, no. 6, pp. 2036-2049, 2023.
[14] O. Ronneberger, P. Fischer, and T. Brox, "U-net: Convolutional networks for biomedical image segmentation," in *Medical image computing and computer-assisted intervention–MICCAI 2015: 18th international conference, Munich, Germany, October 5-9, 2015, proceedings, part III 18*, 2015: Springer, pp. 234-241.
[15] B. Lim, S. Son, H. Kim, S. Nah, and K. Mu Lee, "Enhanced deep residual networks for single image super-resolution," in *Proceedings of the IEEE conference on computer vision and pattern recognition workshops*, 2017, pp. 136-144.
[16] I. Goodfellow *et al.*, "Generative adversarial nets," *Advances in neural information processing systems,* vol. 27, 2014.
[17] J. Liang, J. Cao, G. Sun, K. Zhang, L. Van Gool, and R. Timofte, "Swinir: Image restoration using swin transformer," in *Proceedings of the IEEE/CVF international conference on computer vision*, 2021, pp. 1833-1844.
[18] A. Dosovitskiy, "An image is worth 16x16 words: Transformers for image recognition at scale," *arXiv preprint arXiv:2010.11929,* 2020.
[19] Y.-R. Wang *et al.*, "Low-count whole-body PET/MRI restoration: an evaluation of dose reduction spectrum and five state-of-the-art artificial intelligence models," *European journal of nuclear medicine and molecular imaging,* vol. 50, no. 5, pp. 1337-1350, 2023.
[20] H. Liu *et al.*, "PET image denoising using a deep-learning method for extremely obese patients," *IEEE transactions on radiation and plasma medical sciences,* vol. 6, no. 7, pp. 766-770, 2021.
[21] Q. Liu *et al.*, "A personalized deep learning denoising strategy for low-count PET images," *Physics in Medicine & Biology,* vol. 67, no. 14, p. 145014, 2022.
[22] J.-Y. Zhu, T. Park, P. Isola, and A. A. Efros, "Unpaired image-to-image translation using cycle-consistent adversarial networks," in *Proceedings of the IEEE international conference on computer vision*, 2017, pp. 2223-2232.
[23] L. Zhou, J. D. Schaefferkoetter, I. W. Tham, G. Huang, and J. Yan, "Supervised learning with cyclegan for low-dose FDG PET image denoising," *Medical image analysis,* vol. 65, p. 101770, 2020.
[24] Z. Yuan *et al.*, "QuanTOF: GPU-based List-mode TOF PET Image Reconstruction with Complete Correction Techniques," *Submitted to IEEE transactions on medical imaging,* under review.
[25] A. Rodriguez-Molares *et al.*, "The generalized contrast-to-noise ratio: A formal definition for lesion detectability," *IEEE Transactions on Ultrasonics, Ferroelectrics, and Frequency Control,* vol. 67, no. 4, pp. 745-759, 2019.
[26] Z. Ding, T. Zeng, F. Dong, L. Liu, W. Yang, and T. Long, "An improved PolSAR image speckle reduction algorithm based on structural judgment and hybrid four-component polarimetric decomposition," *IEEE Transactions on Geoscience and Remote Sensing,* vol. 51, no. 8, pp. 4438-4449, 2013.
[27] F. Sattar, L. Floreby, G. Salomonsson, and B. Lovstrom, "Image enhancement based on a nonlinear multiscale method," *IEEE transactions on image processing,* vol. 6, no. 6, pp. 888-895, 1997.
[28] J. Canny, "A computational approach to edge detection," *IEEE Transactions on pattern analysis and machine intelligence,* no. 6, pp. 679-698, 1986.
[29] D. P. Huttenlocher, G. A. Klanderman, and W. J. Rucklidge, "Comparing images using the Hausdorff distance," *IEEE Transactions on pattern analysis and machine intelligence,* vol. 15, no. 9, pp. 850-863, 1993.
[30] K. Zhang, W. Zuo, Y. Chen, D. Meng, and L. Zhang, "Beyond a gaussian denoiser: Residual learning of deep cnn for image denoising," *IEEE transactions on image processing,* vol. 26, no. 7, pp. 3142-3155, 2017.
[31] A. Ghasemabadi, M. Salameh, M. K. Janjua, C. Zhou, F. Sun, and D. Niu, "CascadedGaze: Efficiency in Global Context Extraction for Image Restoration," *arXiv preprint arXiv:2401.15235,* 2024.
[32] Y. Song *et al.*, "Medical ultrasound image quality assessment for autonomous robotic screening," *IEEE Robotics and Automation Letters,* vol. 7, no. 3, pp. 6290-6296, 2022.